\documentclass[twocolumn,showpacs,preprintnumbers,amsmath,amssymb]{revtex4}
\usepackage{graphicx}% Include figure files
\usepackage{dcolumn}% Align table columns on decimal point
\usepackage{bm}% bold math
\begin{document}
%\preprint{APS/123-QED}
\title{Quantum averaging and resonances:\\two-level atom in a one-mode quantized field}% change line with \\
\author{M. Amniat-Talab }
\altaffiliation[Also at ]{Physics Department, Faculty of Science,
Urmia University, P.B. 165, Urmia, Iran.}
\author{S. Gu\'{e}rin}
\author{H.R. Jauslin}
\email{jauslin@u-bourgogne.fr}
 %\homepage{http://www.u-bourgogne.fr/lpub}
\affiliation{Laboratoire de Physique, UMR CNRS 5027,
Universit\'{e} de Bourgogne, B.P. 47870, F-21078 Dijon, France.}
\date{\today}% It is always \today,

\begin{abstract}
We construct a non-perturbative approach based on quantum
averaging combined with resonant transformations to detect the
resonances of a given Hamiltonian and to treat them.  This
approach, that generalizes the rotating-wave approximation, takes
into account the resonances at low field and also at high field
(non-linear resonances). This allows to derive effective
Hamiltonians that contain the qualitative features of the
spectrum, i.e. crossings and avoided crossings, as a function of
the coupling constant. At a second stage the precision of the
spectrum can be improved quantitatively by standard perturbative
methods like contact transformations.  We illustrate this method
to determine the spectrum of a two-level atom interacting with a
single mode of a quantized field.
\end{abstract}
\maketitle

\section{\label{int}Introduction}
Some important features of classical and quantum systems are
determined by resonances of the system  which can not be treated
by perturbative approaches. In the vicinity of resonances the
perturbative formulas display small denominators that lead to the
divergence of the perturbative expansions. A widely used model
that incorporates a one-photon resonance is the Jaynes-Cummings
Hamiltonian extracted from the full dressed Hamiltonian that
describes a two-level system coupled with a single mode of a
quantized field \cite{jaynes}. Its counterpart for an interaction
with a semi-classical laser field is the RWA Hamiltonian
(rotating-wave approximation) \cite{shirley}.

In this article we give a systematic method that allows  to
construct effective Hamiltonians and determine their spectrum  by
treating the resonances with an adaptation of resonant
transformations that were introduced in Ref. \cite{jauslin} in the
context of laser-driven quantum systems in the Floquet
representation. The goal is to obtain the spectrum for a whole
interval of values of a parameter like the coupling constant. This
is needed e.g. in applications where the coupling changes
adiabatically \cite{advchem}, corresponding e.g. to envelopes of
laser pulses or to transversal spatial profiles of cavity fields.
The method is based on the detection of resonances by a projector
derived from Quantum Averaging (QA). We illustrate it on the
problem of a two-level atom interacting with a quantized field and
show that a treatment of all the relevant resonances of the system
in a given range of parameters allows to reproduce with good
accuracy the spectrum of this system.  The treatment of the
resonances yields the qualitative structure of the spectrum -- the
crossings and avoided crossings -- as a function of the coupling
constant. Once this main structure is obtained, one can
systematically improve the quantitative accuracy of the spectrum
by applying perturbative methods. We use contact transformations
with a Kolmogorov-Arnold-Moser (KAM) iteration \cite{jauslin},
that are particularly efficient due to its superconvergent
properties.

The paper is  structured as follows. In Section \ref{prm}, we
describe the method of resonance analysis and  the construction of
effective Hamiltonians. Section \ref{dmpc} contains the
presentation of the model and some preliminary considerations. In
Section \ref{weak}, taking into account the resonances of this
model in the weak-coupling regime, we extract the effective
Hamiltonians by quantum averaging techniques and resonant
transformations. In the weak-coupling regime we have to iterate
this procedure several times to derive the essential structure of
the spectrum  in larger ranges of the coupling constant. In
Sec.\ref{strong} we extract the effective Hamiltonians in the
strong-coupling regime where the qualitative properties of the
spectrum can be globally obtained by some preliminary unitary
transformations and one resonant transformation which treats the
zero-field resonances. We obtain an accurate approximation valid
for all values of the coupling constant that contains all the
qualitative structure. Finally, in Sec.\ref{conc} we give some
conclusions.
\section{\label{prm}principle of the method}
We consider a Hamiltonian $H=H_{0}+\epsilon V$ where $H_{0}$ is
the unperturbed Hamiltonian, $\epsilon V$ is the perturbation and
$\epsilon$ is an ordering parameter. The first analysis of this
problem is in terms of perturbation theory: we look for a KAM-type
unitary transformation $e^{\epsilon W}$ close to the identity that
allows to reduce the order of the perturbation from $\epsilon$ to
$\epsilon^{2}$:
\begin{equation}\label{D}
    e^{-\epsilon W}He^{\epsilon W}=H_{0}+\epsilon D+\epsilon^{2}V_{2}.
\end{equation}
$\epsilon D$ is a remaining term of order $\epsilon$ that
satisfies $\left[H_{0},D\right]=0$. The unknown $W$ and $D$ are
solutions of the following equations \cite{bellissard,jauslin}
\begin{subequations}\label{cohom}
\begin{eqnarray}
\left[H_{0},W\right]+V&=&D,
\\
\left[H_{0},D\right]&=&0.
\end{eqnarray}
\end{subequations}
The remaining perturbation of order $\epsilon^{2}$ is given by
\begin{equation}\label{eps2W}
\epsilon^{2}V_{2}=\sum_{m=2}^{\infty}\frac{\epsilon^{m}}{m!}~\left((m-1)
L^{m-1}_{W}V+L^{m-1}_{W}D\right),
\end{equation}
where $L_{W}$ is defined as
\begin{equation}\label{supop}
L_{W}B=\left[B,W\right].
\end{equation}
 The solutions of Eqs.(\ref{cohom}) can be written in terms of averaging \cite{primas,jauslin}:
\begin{subequations}\label{Vbar}
\begin{eqnarray}
D&=&\overline{V}
\equiv\Pi_{H_{0}}V:=\lim_{\tau\to\infty}\frac{1}{\tau}\int_{0}^{\tau}ds
e^{-iH_{0}s}Ve^{iH_{0}s}
\nonumber\\&=&\sum_{\nu,j,j'}|\nu,j\rangle\langle\nu,j|V|\nu,j'\rangle\langle\nu,j'|,
\\
W&=&\lim_{\tau\to\infty}\frac{-i}{\tau}\int_{0}^{\tau}ds\int_{0}^{s}ds'~e^{-iH_{0}s'}(V-\Pi_{H_{0}}V)e^{iH_{0}s'}
\nonumber\\&=&-\sum_{\nu,j,j',\nu'\neq\nu}\frac{|\nu,j\rangle\langle\nu,j|V|\nu',j'\rangle\langle\nu',j'|}{E_{\nu}^{(0)}-E_{\nu'}^{(0)}},
\end{eqnarray}
\end{subequations}
 where $\nu$ labels the different eigenvalues $E_{\nu}^{(0)}$ of
$H_{0}$, and $j$ is a degeneracy index which distinguishes
different basis vectors $|\nu,j\rangle$ of the degeneracy
eigenspace. The operator $\Pi_{H_{0}}$ is the projector on the
kernel of the application $A \mapsto [H_{0},A]$. We remark that
the integral representation of $D,W$ in Eqs. (\ref{Vbar}) can be
also well-defined in  cases where $H_{0}$ has a continuum
spectrum. The units are chosen  such that $\hbar=1$. In the
following discussion, we do not write explicitly the ordering
parameter $\epsilon$.

A \emph{resonance} is defined as a degeneracy of an eigenvalue
$E_{\nu}^{(0)}$ of $H_{0}$ and  is said to be \emph{active} if the
perturbation V has nonzero matrix elements in the degeneracy
subspace of $E_{\nu}^{(0)}$: $\langle\nu,j|V|\nu,j'\rangle \neq
0$~~for some $j,j'$. Otherwise the resonance is called
\emph{passive} or \emph{mute}. An active resonance renders $W$
arbitrarily large close to the degeneracy and makes the
perturbative expansion diverge. The method we present here, is a
construction designed to avoid such divergences. We remark that
the concept of resonance is defined intrinsically for $H_{0}$,
while the distinction between active and passive depends on the
relation between $H_{0}$ and $V$. The analysis of the resonances
involves thus three aspects:
\begin{itemize}
    \item Decomposition of the Hamiltonian into $H=H_{0}+V$. Different decompositions can be considered for different
regimes of the parameters of $H$.
    \item Determination of degenerate eigenvalues of $H_{0}$.
    \item Detection of the \emph{resonant terms} in the perturbation $V$
that couple these degenerate eigenstates.
\end{itemize}
 The resonant terms of $V$ can be detected by
projectors of type $\Pi_{H_{0}}$ that extract a block-diagonal
part of $V$ relative to $H_{0}$, where the blocks are generated by
the degeneracy subspaces. In absence of active resonances, when
all the eigenvalues of $H_{0}$ are non-degenerate or when the
resonances are mute, the matrix representation of $\Pi_{H_{0}}V$
is in fact diagonal in the eigenbasis of $H_{0}$. In presence of
active resonances, the block-diagonal effective Hamiltonian that
takes into account the considered resonance of the original
Hamiltonian can be written as
\begin{equation}\label{Heff}
    H^{\text{eff}}=H_{0}+\Pi_{H_{0}}V.
\end{equation}
We will call the transformation that diagonalizes $H^{\text{eff}}$
\emph{Resonant Transformation} (RT). The Hamiltonian
$H=H^{\text{eff}}+(V-\Pi_{H_{0}}V)$ is transformed under RT
(denoted $\mathcal{R}$) as follows:
\begin{eqnarray}\label{h1}
    H_{1}=\mathcal{R}^{\dag}H\mathcal{R}&=&\mathcal{R}^{\dag}H^{\text{eff}}\mathcal{R}+\mathcal{R}
    ^{\dag}(V-\Pi_{H_{0}}V)\mathcal{R}\nonumber\\
    &=:&H_{1}^{(0)}+V_{1},
\end{eqnarray}
where $H_{1}^{(0)}$ is defined as the new renormalized reference
Hamiltonian and $V_{1}$ is the new perturbation. If
$H_{1}^{(0)}+V_{1}$ does not have any other active resonance in
the considered range of the coupling constant, we can at a second
stage improve the spectrum by a KAM-type perturbative expansion
which is expected to converge. If there are other active
resonances, we have to iterate the renormalization procedure by
applying another RT. We remark that there are cases of
multi-photon resonances where the resonant terms appear only after
applying one or several contact transformations.
\section{\label{dmpc}description of the model and preliminary considerations}
We consider as an illustration a two-level atom interacting with a
single mode of a quantized  field described by
\begin{equation}\label{tls}
    H=\omega (a^{\dag}a+1/2)\otimes\openone_{2}+\frac{\omega_{0}}{2}\openone\otimes\sigma_{z}
    +g(a+a^{\dag})\otimes\sigma_{x},
\end{equation}
 where $a$, $a^{\dag}$ are the annihilation and
creation operators for the field mode with the commutation
relation $[a,
a^{\dag}]=\openone=\sum_{n=0}^{\infty}|n\rangle\langle n|$ ,
$\sigma_{z},\sigma_{x}$ are Pauli matrices and $\openone_{2}$ is
the $2\times2$ identity matrix. Here $\omega$ is the frequency of
field mode,
 $\omega_{0}$ is the energy difference of the two atomic states and $g$ is the
 dipole-coupling  between the field mode and the atom.
 This Hamiltonian acts on
the  Hilbert space $\mathcal{K}= \mathcal{F}\otimes\mathcal{H}$
where $\mathcal{H}=\mathbb{C}^{2}$ is the Hilbert space of the
atom generated by $|\pm\rangle$ (eigenvectors of $\sigma_{z}$) and
$\mathcal{F}$ is the Fock space of the field mode generated by the
orthonormal basis $\{|n\rangle~;~n=0,1,2,\cdots\}$, $n$ being the
photon number of the field.

For this system there is a parity operator
\begin{equation}\label{pari}
P=e^{i\pi a^{\dag}a}\otimes
    \sigma_{z}=\sum_{n=0}^{\infty}(-1)^{n}|n\rangle\langle
    n|\otimes\sigma_{z},
\end{equation}
with the properties
\begin{equation}\label{pariprop}
    [P,H]=0,~~~~P=P^{\dag},~~~~P^{2}=\openone_{\mathcal{K}}\equiv\openone\otimes
    \openone_{2}.
\end{equation}
As a consequence, the eigenstates of $H$ can be separated into two
symmetry classes, even or odd, under $P$:
\begin{equation}\label{parieig}
    P|\phi_{n,\pm}\rangle=\pm|\phi_{n,\pm}\rangle,~~~~H|\phi_{n,\pm}\rangle=E_{n,\pm}|\phi_{n,\pm}\rangle.
\end{equation}
The parity operator also commutes with any operator that depends
only on $N=a^{\dag}a$ and $\sigma_{z}$.

  In spite of the simple form of (\ref{tls}), its exact
  solutions are not known. This can be related to the fact that the classical
  limit of this model is non-integrable \cite{miloni}. This model is of great interest  as a physical model in
quantum optics \cite{allen,lais,cohen,frasca} and quantum chaos
\cite{graham1,graham2}. Some approximate solutions of this model
have been studied among many others in \cite{franchok,tur1} using
different formalisms.

 The conceptual framework for the solution of this system based on the
construction of unitary transformations can be described as
follows: First, we decompose the Hamiltonian in two terms
 as $H=H_{0}+V$. Depending on the considered ranges of the parameters of the system, different
 decompositions may be considered. $H_{0}$ is \emph{a priori} an operator that is a regular function exclusively
of the operators $N$ and $\sigma_{z}$. The operators $N$ and
$\sigma_{z}$ can be considered in the present model as quantum
analogues of classical \emph{global actions} \cite{weigert}, and
$H_{0}$ can be labelled \emph{integrable}. The perturbation $V$
contains functions that involve also the other operators
$a,a^{\dag},\sigma_{x},\sigma_{y}$. The goal is to determine a
unitary transformation $U$, that should be expressed in terms of
well-behaved regular functions of
$a,a^{\dag},\sigma_{x},\sigma_{y},\sigma_{z}$, such that:
\begin{equation}\label{S}
    U^{\dag}\left(H_{0}(N,\sigma_{z})+V(a,a^{\dag},\sigma_{x},\sigma_{y},\sigma_{z})\right)U=H'(N,\sigma_{z}),
\end{equation}
where $H'$ is a regular function $f$ exclusively of the action
operators $N,\sigma_{z}$: $H'(N,\sigma_{z})= f(N,\sigma_{z})$.
With this transformation the eigenvectors of $H$ can be expressed
as $|\phi_{n,\pm}\rangle=U(|n\rangle\otimes|\pm\rangle)$ and the
corresponding eigenvalues as $E_{n,\pm}=f(n,\pm1)$ where
$N|n\rangle=n|n\rangle$ and
$\sigma_{z}|\pm\rangle=\pm|\pm\rangle$.

We remark that in our context the important property for singling
out the operators $N,\sigma_{z}$ is that they commute with each
other and their spectrum and eigenvectors are explicitly
available. The question of whether for a given model there exists
a regular unitary transformation $U$ that accomplishes the above
requirement is, to our knowledge, an open problem.

Most of the perturbative approaches can be interpreted as methods
to find approximations of the transformation $U$. The presence of
resonances is one of the central difficulties in the construction
of $U$, as will be made precise below. In this paper we discuss an
iterative approach that consists of constructing first some
approximations of $U$ that take into account the dominating
effects of a certain number of resonances. The transformations
involved in this stage are far from the identity and have a
clearly non-perturbative character. Once we have a transformation
that takes into account the main effect of a set of resonances
that are relevant in a considered interval of the coupling
constant $g$, a perturbative approach (like the KAM, Van Vleck, or
other types of the contact transformation) can be applied to
improve the approximation quantitatively. The transformations
involved in this second stage can be considered as deformations of
the identity, since they can be written in the form $e^{W}$. This
stage cannot be implemented if the resonances are not taken care
of beforehand. Indeed the perturbative formulations diverge close
to resonances due to the appearance of \emph{small denominators }
as can be seen in Eq. (\ref{Vbar}-b).

As in classical mechanics, the construction of the transformation
$U$ leading to a Hamiltonian that contains only  action variables
can often be considered in two steps: $U=U_{1}U_{2}$. In the first
step, that is called \emph{reduction}, the Hamiltonian is
transformed by $U_{1}$ into a form that contains functions of
$\sigma_{z},\sigma_{x},\sigma_{y}$ and $N$, but not of $a$ and
$a^{\dag}$. The degree of freedom of the field is made trivial and
the number of non-trivial degrees of freedom is thus reduced by
one. When we apply this reduction to the effective Hamiltonian
(\ref{Heff}), we obtain a \emph{reduced effective Hamiltonian}. We
remark that in the literature, this ``reduced effective
Hamiltonian" is often called simply ``effective Hamiltonian".
%Generally in the literature, it is this reduced Hamiltonian
%$U_{1}^{\dag}HU_{1}$ which is called \emph{effective Hamiltonian}.
In the second step, the reduced Hamiltonian is transformed under
$U_{2}$ into a form that contains functions of only $N$ and
$\sigma_{z}$. For the  model (\ref{tls}), the reduction step
corresponds to diagonalization in the Fock space and the second
step corresponds to diagonalization in the atomic Hilbert space
which in this case is trivial. The construction of the RT is based
on this reduction procedure.
\section{\label{weak}effective Hamiltonians in the weak-coupling regime}
In this section we consider the Hamiltonian (\ref{tls}) at
resonance $\omega_{0}=\omega$ in the weak coupling regime, so that
$H$ can be decomposed as follows:
\begin{eqnarray}\label{rwt1}
 H&=&H_{0}+V, \nonumber\\
    H_{0}(N,\sigma_{z})&=&\omega (N+1/2)\otimes\openone_{2}+\frac{\omega_{0}}{2}\openone
    \otimes\sigma_{z},\nonumber\\
    V(a,a^{\dag},\sigma_{x},g)&=&
    g(a+a^{\dag})\otimes \sigma_{x}.
\end{eqnarray}
 The eigenvalues and eigenvectors of $H_{0}$ are:
\begin{eqnarray}\label{rwt2}
    E_{n,\pm}^{(0)} &=& \omega(n+1/2)\pm\omega_{0}/2,\nonumber\\
    |\phi_{n,\pm}^{(0)}\rangle &=& |n,\pm\rangle=
    |n\rangle\otimes|\pm\rangle ,\nonumber\\
    |n,+\rangle &=& \left(%
\begin{array}{c}
  |n\rangle \\
  0 \\
\end{array}%
\right), ~~~~|n,-\rangle=\left(%
\begin{array}{c}
  0 \\
  |n\rangle \\
\end{array}%
\right).
\end{eqnarray}
 For $\omega_{0}=\omega$ there is a one
photon resonance which corresponds to the degeneracies
$E_{n,+}^{(0)}=E_{n+1,-}^{(0)}$. The degeneracy eigenspaces are
spanned by the vectors $|\phi_{n,+}^{(0)}\rangle$ and
$|\phi_{n+1,-}^{(0)}\rangle$. The resonant part of $V$ is obtained
by (\ref{Vbar}-a):
\begin{eqnarray}\label{rwt3}
    V_{res}:=\Pi_{H_{0}}V&=&\sum_{n=0}^{\infty}\big(|n,+\rangle\langle n,+|V|n+1,-\rangle\langle n+1,-|\nonumber\\
    &+&
    |n+1,-\rangle\langle n+1,-|V|n,+\rangle\langle n,+|\big)\nonumber\\
    &=&g\left(%
\begin{array}{cc}
  0 & a \\
  a^{\dag} & 0 \\
\end{array}%
\right),
\end{eqnarray}
 where we have used the relations
\begin{equation}\label{rwt5}
    a=\sum_{n=0}^{\infty}\sqrt{n+1}~|n\rangle\langle n+1|, ~~~
     ~~~ a^{\dag}=\sum_{n=0}^{\infty}\sqrt{n+1}~|n+1\rangle\langle
    n|.
\end{equation}
The effective Hamiltonian  containing the one-photon resonance is
the so-called Jaynes-Cummings Hamiltonian that can be written as
\begin{eqnarray}\label{jcm}
    H_{0}^{\text{eff}}&=&H_{JC}=H_{0}+\Pi_{H_{0}}V=\omega
    (N+1/2)\otimes\openone_{2}\nonumber\\&+&\frac{\omega}{2}\openone\otimes\sigma_{z}+g\left(%
\begin{array}{cc}
  0 & a \\
  a^{\dag} & 0 \\
\end{array}%
\right).
\end{eqnarray}
$H_{JC}$ is a good approximation of (\ref{tls}) for low energies
in the limit $g\ll\omega_{0},|~\omega-\omega_{0}|\ll\omega_{0}$.
In this limit, the so-called counter-rotating terms $g\left(%
\begin{array}{cc}
  0 & a^{\dag} \\
  a & 0 \\
\end{array}%
\right)$ can be discarded (rotating-wave approximation). $H$ can
thus be written as
$H=H_{0}^{\text{eff}}(N,a,a^{\dag},\sigma_{x},\sigma_{y};g)+(V-\Pi_{H_{0}}V)$.
Next we  transform  $H_{0}^{\text{eff}}$ by a resonant
transformation $\mathcal{R}_{1}$ to a regular function of
exclusively the action operators $N,\sigma_{z}$. Every resonant
transformation is performed in two steps. To diagonalize
$H_{0}^{\text{eff}}$ in the Fock space (the reduction step of the
RT denoted  $R_{1}$) we define a transformation in such a way that
the following condition is satisfied:
\begin{equation}\label{rwt6}
R_{1}^{\dag}V_{res}R_{1}= f(N)\otimes\sigma_{x},
\end{equation}
where $f$ is a regular function of $N$ which has to be determined.
We require furthermore that $R_{1}^{\dag}H_{0}R_{1}$ stays a
function of only $N$ and $\sigma_{z}$. A suitable transformation
satisfying these conditions is
 \begin{equation}\label{rwt7}
    R_{1}:=\left(%
\begin{array}{cc}
  (aa^{\dag})^{-1/2}a &~~ 0 \\
  0 &~~ \openone \\
\end{array}%
\right)\equiv\left(%
\begin{array}{cc}
  \sum_{n=0}^{\infty}|n\rangle\langle n+1| &~~ 0 \\
  0 &~~ \openone \\
\end{array}%
\right).
\end{equation}
 This transformation is not unitary but \emph{isometric} \cite{reed}:
\begin{equation}\label{rwt9}
R_{1}R_{1}^{\dag}=\openone_{\mathcal{K}}~~,~~R_{1}^{\dag}R_{1}=\openone_{\mathcal{K}}-\left(%
\begin{array}{cc}
  |0\rangle\langle0| &~~ 0 \\
  0 &~~ 0 \\
\end{array}%
\right),
\end{equation}
where we have used the identity
$a^{\dag}(N+1)^{-1}a=\openone-|0\rangle\langle0|$. Applying this
transformation on the resonant term gives
\begin{equation}\label{rwt8}
R_{1}^{\dag}V_{res}R_{1}=g~a^{\dag}(aa^{\dag})^{-1/2}a\otimes\sigma_{x}=g\sqrt{N}\otimes\sigma_{x}
\end{equation}
and $H$ is transformed under $R_{1}$ as
\begin{equation}\label{H1}
    H_{R_{1}}=R_{1}^{\dag}HR_{1}=\omega N\otimes
    \openone_{2}+g\sqrt{N}\otimes \sigma_{x}+g\left(%
\begin{array}{cc}
  0 &~~ A^{\dag} \\
  A &~~ 0 \\
\end{array}%
\right),
\end{equation}
 where
\begin{equation}\label{A}
    A=a(aa^{\dag})^{-1/2}a=\sum_{n=0}^{\infty}\sqrt{n+1}~|n\rangle\langle
    n+2|,
\end{equation}
with the properties:
\begin{equation}\label{Aprop}
    AA^{\dag}=aa^{\dag}~~,~~A^{\dag}A=a^{\dag}a-\openone+|0\rangle\langle0|.
\end{equation}
To each eigenvector $|\phi\rangle$ of $H$ corresponds an
eigenvector $R_{1}^{\dag}|\phi\rangle$ of $H_{R_{1}}$, since:
\begin{equation}\label{HR1}
    H_{R_{1}}R_{1}^{\dag}|\phi\rangle=R_{1}^{\dag}HR_{1}R_{1}^{\dag}|\phi\rangle=\lambda
    R_{1}^{\dag}|\phi\rangle.
\end{equation}
We remark that $R_{1}^{\dag}|\phi\rangle\neq 0 ~~\forall
|\phi\rangle \in \mathcal{K}$. Every eigenvalue of the original
Hamiltonian $H$ is also an eigenvalue of the transformed
Hamiltonian $H_{R_{1}}$. However since $R_{1}|0,+\rangle=0$, there
is a difference in the spectrum between $H$ and $H_{R_{1}}$ :
$H_{R_{1}}$ has an extra zero eigenvalue with eigenvector
$|0,+\rangle$. The spurious eigenvalue can be detected and
eliminated after applying the transformation. Indeed, since
$|0,+\rangle$ is not coupled to any vector in its orthogonal
complement, one can eliminate it from the rest of the calculation
by taking the projection of $H_{R_{1}}$ into the orthogonal
complement
    $H_{R_{1},\bot(0,+)}=P_{\bot(0,+)}H_{R_{1}}P_{\bot(0,+)}$ with
    $P_{\bot(0,+)}=\openone_{\mathcal{K}}-|0,+\rangle\langle0,+|$.This difference between
unitary and isometric transformations was not taken into account
in \cite{hussin} in diagonalizing the Jaynes-Cummings Hamiltonian.

The second step of the RT is the diagonalization of
$R_{1}^{\dag}H_{0}^{\text{eff}}R_{1}=\omega
N\otimes\openone_{2}+\sqrt{N}\otimes\sigma_{x}$ in the atomic
Hilbert space. This can be performed by a $\pi/2$ rotation around
the y-axis:
\begin{equation}\label{T}
    T=e^{-i\frac{\pi}{4}\sigma_{y}}=\frac{1}{\sqrt{2}}\left(%
\begin{array}{cc}
  1 & -1 \\
  1 & 1 \\
\end{array}%
\right),
\end{equation}
with the properties
\begin{equation}\label{Tp}
   T^{\dag}\sigma_{x}T=\sigma_{z},~~~~T^{\dag}\sigma_{z}T=-\sigma_{x}.
\end{equation}
%%%%%%%%%%%%%%%%%%%%%%%%%%%%%%%%%%%%%%%%%%%%%%%%%%%%%%%%%%%%%%%%%%%%%%%%
\begin{figure}
  \includegraphics[width=8.5cm]{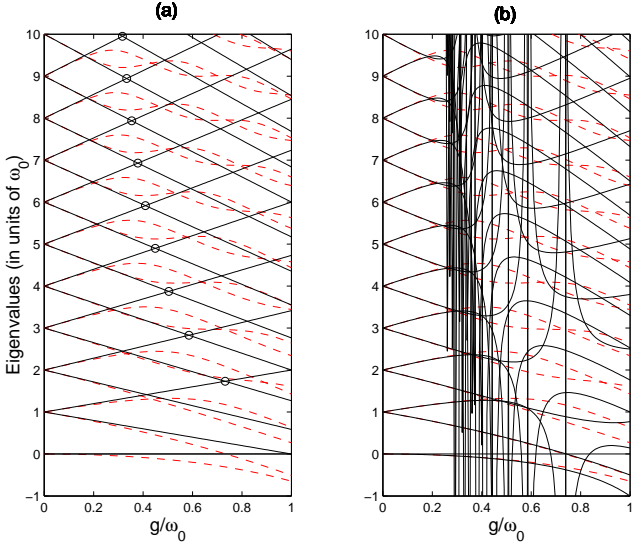}\\
  \caption{Comparison of exact numerical eigenvalues (dashed lines) of
  (\ref{tls}) for one-photon resonance $\omega=\omega_{0}$ with the approximate ones (solid lines) obtained
   after (a) 1 one-photon RT
 given by (\ref{Ejcm}), (b) 1 one-photon RT plus 1 iteration of KAM-type perturbative
 expansion. The divergence observed around $g/\omega_{0}=0.3$ in panel (b) is due to the active nonlinear resonances
 of $H_{1}^{(0)}$ occurred at the degeneracies marked by circles in panel (a). One can see clearly that
 the locations of these resonances depend on $n$ according to Eq. (\ref{gn}).}\label{1rt+kam}
\end{figure}
%%%%%%%%%%%%%%%%%%%%%%%%%%%%%%%%%%%%%%%%%%%%%%%%%%%%%%%%%%%%%%%%%%%%
However, since  the spurious eigenvector $|0,+\rangle$ can be
separated and $|0,-\rangle$ is already an eigenvector of
$R_{1}^{\dag}H_{0}^{\text{eff}}R_{1}$, the transformation $T$ must
be applied only on the subspace with $n\geq1$ photons. The
complete transformation (denoted $T_{1}$) reads thus
\begin{equation}\label{T1}
T_{1}=P_{0}\otimes \openone_{2}+P_{\bot0}\otimes T,
\end{equation}
where
\begin{equation}\label{P0}
P_{0}=|0\rangle\langle0|,~~~~P_{\bot0}=\sum_{n=1}^{\infty}|n\rangle\langle
n|.
\end{equation}
 Applying $T_{1}$ gives
\begin{eqnarray}\label{secRT}
    H_{1}&:=&T_{1}^{\dag}R_{1}^{\dag}HR_{1}T_{1}\nonumber\\
    &=&H_{1}^{(0)}(N,\sigma_{z};g)+V_{1}(a,a^{\dag},\sigma_{z},\sigma_{x},\sigma_{y};g),
\end{eqnarray}
with
\begin{eqnarray}
    H_{1}^{(0)}&=&\omega N\otimes
    \openone_{2}+g\sqrt{N}\otimes \sigma_{z},\nonumber\\
    V_{1}&=&g/2\left(%
\begin{array}{cc}
  A_{\bot0}+A_{\bot0}^{\dag} &~~ -A_{\bot0}+A_{\bot0}^{\dag} \\
  A_{\bot0}-A_{\bot0}^{\dag} &~~ -A_{\bot0}-A_{\bot0}^{\dag} \\
\end{array}%
\right)\nonumber\\
&+&\frac{g}{\sqrt{2}}\left(%
\begin{array}{cc}
  0 &~~ |2\rangle\langle 0| \\
  |0\rangle\langle2| &~~ -|2\rangle\langle0|-|0\rangle\langle2| \\
\end{array}%
\right),
\end{eqnarray}
where
\begin{equation}\label{Abot0}
    A_{\bot0}=P_{\bot0}AP_{\bot0}=\sum_{n=1}^{\infty}\sqrt{n+1}~|n\rangle\langle
    n+2|,
\end{equation}
 and use has been made of the relations
\begin{equation}\label{Abot}
    AP_{0}=P_{0}A^{\dag}=0,~~~~P_{0}AP_{\bot0}=|0\rangle\langle2|.
\end{equation}
The first RT, is thus the combination of $R_{1}T_{1}$. Since the
transformation $R_{1}$ dresses the upper atomic state by (--1)
photon \cite{cohen}, $\mathcal{R}_{1}=R_{1}T_{1}$ can be called a
one-photon RT.

$H_{1}^{(0)}$ is in fact the diagonalized Jaynes-Cummings
Hamiltonian in the resonant case with the eigenvalues
\begin{equation}\label{Ejcm}
E_{1,(n,\pm)}^{(0)}(g)=\omega n \pm g\sqrt{n},~~~~n=0,1,2,\cdots.
\end{equation}
The eigenvalues and therefore the degeneracies of $H_{1}^{(0)}$
depend on the coupling constant $g$. For small enough $g$ and low
energies, $H_{1}^{(0)}$ does not have other degeneracies besides
the ones at $g=0$ for which the new perturbation $V_{1}$ does not
have resonant terms, and we can apply KAM-type transformations to
improve quantitatively the precision of the spectrum  by
iteration. A single KAM transformation (which is essentially
equivalent to second order perturbation theory) already gives
quite good precision, as shown in Fig. (\ref{1rt+kam}-b) for
$g/\omega_{0}<0.25$ for energies smaller than $10\omega_{0}$. If
we take large enough $g$ or larger energies, we encounter new
resonances which appear at some specific finite values of $g$.
These resonances are called \emph{field-induced resonances} or
\emph{nonlinear resonances}. For larger values of the coupling
($g/\omega_{0}\approx 0.3$ for the shown energy interval in Fig.
(\ref{1rt+kam}-b)), where we encounter nonlinear resonances, the
KAM iteration diverges. The eigenvalues of $H_{1}^{(0)}$ are
degenerate at $g_{n}=\omega/(\sqrt{n}+\sqrt{n+1})$
 as $E_{1,(n,+)}^{(0)}(g_{n})=E_{1,(n+1,-)}^{(0)}(g_{n})$. But the corresponding
resonant terms in $V_{1}$ are zero due to parity (mute
resonances). The next degeneracies appear at
\begin{equation}\label{gn}
    g_{n}=2\omega/(\sqrt{n}+\sqrt{n+2}),
\end{equation}
as
\begin{equation}\label{Edeg}
    E_{1,(n,+)}^{(0)}(g_{n})=E_{1,(n+2,-)}^{(0)}(g_{n}),
\end{equation}
which have been marked by circles in figure (\ref{1rt+kam}-a). All
the other resonances are mute.  There is an infinite family of
nonlinear resonances located at different values of the coupling
$g_{n}$. We observe from (\ref{gn}) that for higher energies the
nonlinear resonances appear for arbitrary small coupling
($\lim_{n\rightarrow\infty} g_{n}=0$). We can extract the resonant
terms corresponding to the whole family in a single step by
working with the combined projector $\sum_{n}
\Pi_{H_{1}^{(0)}(g_{n})}$. The resonant terms in $V_{1}$
corresponding to the degeneracies (\ref{Edeg}) are
 \begin{eqnarray}\label{res2}
    \sum_{n}\Pi_{H_{1}^{(0)}(g_{n})}V_{1}&=&-\frac{g}{2}\left(%
\begin{array}{cc}
  0 & A_{\bot0} \\
  A^{\dag}_{\bot0} & 0 \\
\end{array}%
\right)\nonumber\\
&-&\frac{g}{\sqrt{2}}\left(%
\begin{array}{cc}
  0 & 0\\
  0 & |2\rangle\langle0|+|0\rangle\langle2| \\
\end{array}%
\right),
\end{eqnarray}
and the new effective Hamiltonian is thus
\begin{equation}\label{h1eff}
    H_{1}^{\text{eff}}=\omega N\otimes
    \openone_{2}+g\sqrt{N}\otimes
    \sigma_{z}+\sum_{n}\Pi_{H_{1}^{(0)}(g_{n})}V_{1}.
\end{equation}
To diagonalize $H_{1}^{\text{eff}}$, it can be decomposed
according to three orthogonal subspaces:
\begin{eqnarray}\label{H1sep}
    H_{1}^{\text{eff}}&=&P_{(0,2,-)}H_{1}^{\text{eff}}P_{(0,2,-)}+P_{(0,+)}H_{1}^{\text{eff}}
    P_{(0,+)}+P_{\bot}H_{1}^{\text{eff}}P_{\bot}\nonumber\\
    &=&H_{1}^{\text{eff}}P_{(0,2,-)}+H_{1}^{\text{eff}}P_{(0,+)}+H_{1}^{\text{eff}}P_{\bot}.
\end{eqnarray}
where the projectors, which commute with $H_{1}^{\text{eff}}$, are
defined by
\begin{widetext}
\begin{eqnarray}\label{Prs}
P_{(0,2,-)}&=&\left(%
\begin{array}{cc}
  0 & 0 \\
  0 & |0\rangle\langle0|+|2\rangle\langle2| \\
\end{array}%
\right),~~~~~~~~~~~~~~
P_{(0,+)}=\left(%
\begin{array}{cc}
  |0\rangle\langle0| & 0 \\
  0 & 0 \\
\end{array}%
\right),\nonumber\\
P_{\bot}&=&\openone_{\mathcal{K}}-P_{(0,2,-)}-P_{(0,+)}
=\left(%
\begin{array}{cc}
  \sum_{n=1}^{\infty}|n\rangle\langle n| & 0 \\
  0 & \sum_{n=1,\neq2}^{\infty}|n\rangle\langle n| \\
\end{array}%
\right).
\end{eqnarray}
which leads to
\begin{eqnarray}
  H_{1}^{\text{eff}}P_{(0,+)}&=&0,~~~~~~~~
   H_{1}^{\text{eff}}P_{(0,2,-)}= \left[(2\omega-g\sqrt{2})|2\rangle\langle2|
   -\frac{g}{\sqrt{2}}(|2\rangle\langle0|+|0\rangle\langle2|)\right]\left(%
\begin{array}{cc}
  0 & 0 \\
  0 & 1 \\
\end{array}%
\right), \nonumber\\
  H_{1}^{\text{eff}}P_{\bot} &=&\omega\left(%
\begin{array}{cc}
  \sum_{n=1}^{\infty}n|n\rangle\langle n| & 0 \\
  0 & +\sum_{n=1,n\neq2}^{\infty}n|n\rangle\langle n| \\
\end{array}%
\right)
+g\left(%
\begin{array}{cc}
  \sum_{n=1}^{\infty}\sqrt{n}~|n\rangle\langle n| & 0 \\
  0 & \sum_{n=1,n\neq2}^{\infty}\sqrt{n}~|n\rangle\langle n| \\
\end{array}%
\right)\nonumber\\
&-&\frac{g}{2}\left(%
\begin{array}{cc}
  0 & A_{\bot0} \\
  A^{\dag}_{\bot0} & 0 \\
\end{array}%
\right).
\end{eqnarray}
 $H_{1}^{\text{eff}}P_{(0,2,-)}$ can be directly diagonalized
by
\begin{equation}\label{S02}
    R_{(0,2,-)}=P_{(0,2,-)}\left(%
\begin{array}{cc}
  0 &~~~~ 0 \\
  0 & ~~~~\cos\theta\big(|2\rangle\langle2|-|0\rangle\langle0|\big)-\sin\theta\big(|2\rangle\langle0|+|0\rangle\langle2|\big) \\
\end{array}%
\right)P_{(0,2,-)},
\end{equation}
where the angle $\theta$ is defined by the relation
\begin{equation}\label{teta}
    \tan2\theta=\frac{g\sqrt{2}}{2\omega-g\sqrt{2}},~~~~~~~~0\leq\theta<\frac{\pi}{2}.
\end{equation}
and the corresponding eigenvalues are
\begin{equation}\label{E1eff}
  E_{1,(0,+)}^{\text{eff}}=0,~~~~~~~~E_{1,(n=0,2,-)}^{\text{eff}} = \omega
  -\frac{g}{\sqrt{2}}\pm\frac{1}{2}\sqrt{(2\omega-g\sqrt{2})^{2}+2g^{2}}.
\end{equation}

 The reduction step of the second RT to diagonalize  $H_{1}^{\text{eff}}P_{\bot}$ in the Fock space can be
defined as
\begin{equation}\label{R2}
    R_{2,\bot}:=P_{\bot}\left(%
\begin{array}{cc}
  (A_{\bot0}A_{\bot0}^{\dag})^{-1/2}A_{\bot0} &~~ 0 \\
  0 &~~ \openone\\
\end{array}%
\right)P_{\bot}
=\left(%
\begin{array}{cc}
  \sum_{n=1}^{\infty}|n\rangle\langle n+2|~~ & ~0 \\
  0 &  \sum_{n=1,\neq2}^{\infty}|n\rangle\langle n|\\
\end{array}%
\right)
\end{equation}
with the properties
\begin{equation}
R_{2,\bot}R_{2,\bot}^{\dag}=P_{\bot},~~~~~~R_{2,\bot}^{\dag}R_{2,\bot}=P_{\bot}-\left(%
\begin{array}{cc}
  |1\rangle\langle1|+|2\rangle\langle2| &~ 0 \\
  0 &~ 0 \\
\end{array}%
\right).
\end{equation}
Equation (\ref{R2}) shows that  $R_{2,\bot}$ dresses the upper
atomic state by (--2) photons. Therefore $\mathcal{R}_{2,\bot}$
can be  called a two-photon RT. Since
$R_{2,\bot}|1,+\rangle=0=R_{2,\bot}|2,+\rangle$, the spectrum of
$R_{2,\bot}^{\dag}H_{1}^{\text{eff}}P_{\bot}R_{2,\bot}$ has two
extra zero eigenvalues relative to the spectrum of
$H_{1}^{\text{eff}}P_{\bot}$. Applying $R_{2,\bot}$ gives
\begin{eqnarray}
  R_{2,\bot}^{\dag}H_{1}^{\text{eff}}P_{\bot}R_{2,\bot} &=& \omega \left(%
\begin{array}{cc}
  \sum_{n=3}^{\infty}(n-2)|n\rangle\langle n| & 0 \\
  0 &  \sum_{n=1,\neq2}^{\infty}n|n\rangle\langle n|\\
\end{array}%
\right)+g\left(%
\begin{array}{cc}
  \sum_{n=3}^{\infty}\sqrt{n-2}~|n\rangle\langle n| & 0 \\
  0 &  -\sum_{n=1,\neq2}^{\infty}\sqrt{n}~|n\rangle\langle n|\\
\end{array}%
\right) \nonumber\\
   &-&g/2\sum_{n=3}^{\infty}\sqrt{n-1}~|n\rangle\langle n|\otimes\sigma_{x}.%\nonumber\\&+&S_{2}^{\dag}~V_{1,non-res}~S_{2}
\label{apR2}\end{eqnarray}
%%%%%%%%%%%%%%%%%%%%%%%%%%%%%%%%%%%%%%%%%%%%%%%%%%%%%%%%%%%%%%%%%%%%%%%%
\begin{figure*}
  \includegraphics[width=16cm]{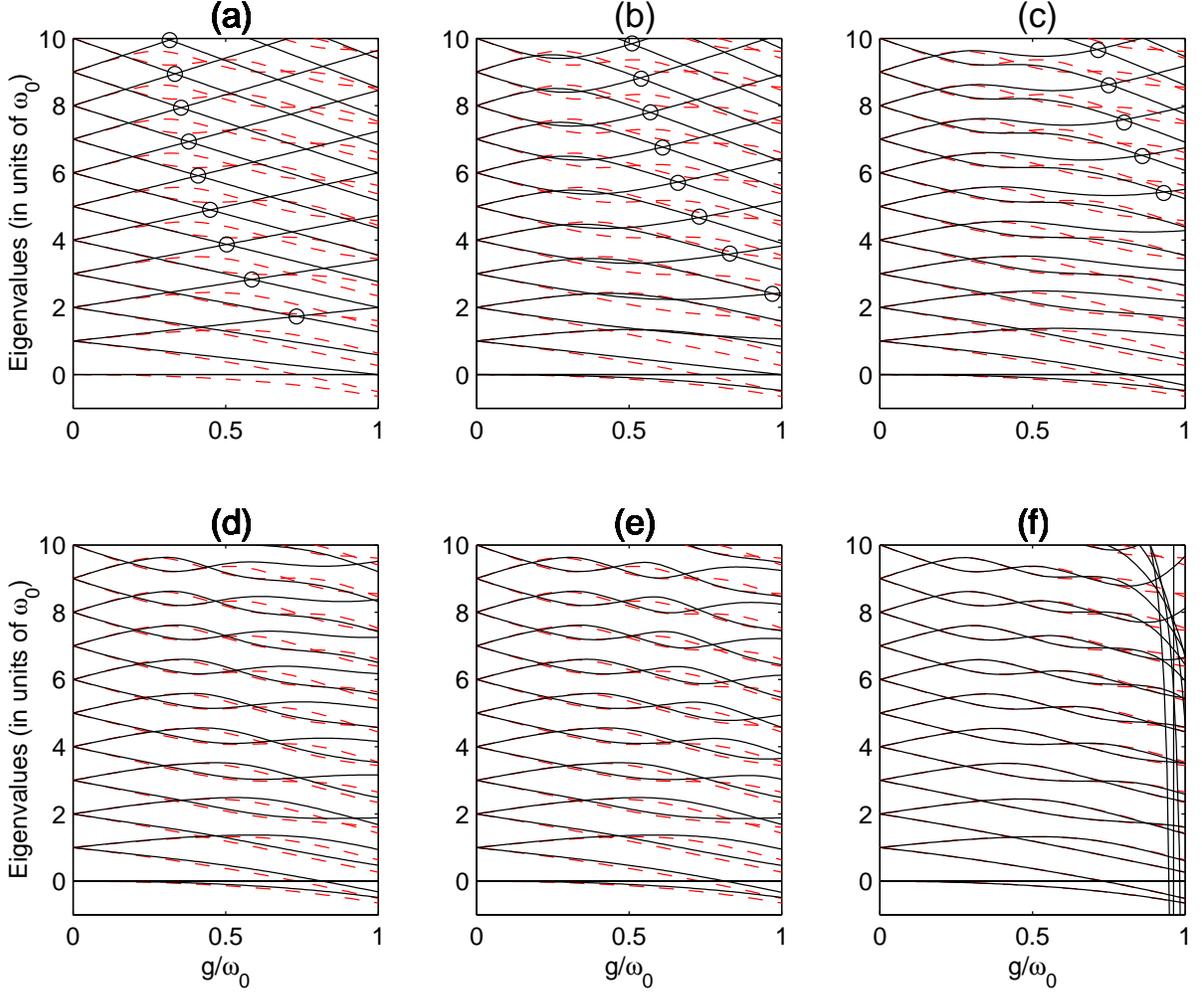}\\
  \caption{Comparison of the exact numerical eigenvalues (dashed lines) of
  (\ref{tls}) for one-photon resonance $\omega=\omega_{0}$ with the approximate ones (solid lines) obtained respectively after (a) 1 one-photon RT
 given by (\ref{Ejcm}), (b) 1 one-photon RT plus 1 two-photon RT  given by (\ref{E2-0}), (c) 1 one-photon
  RT plus 2 two-photon RT, (d) 1 one-photon RT plus 3 two-photon RT, (e) 1 one-photon RT plus 4 two-photon RT,
 (f) 1 one-photon RT plus 4 two-photon RT
  plus 1 iteration of KAM-type perturbative expansion. The divergence of the KAM transformation observed close
  to $g/\omega_{0}=1$ in panel (f) is due to the presence of  active resonances at larger values of $g$.}\label{tlsw}
\end{figure*}
%%%%%%%%%%%%%%%%%%%%%%%%%%%%%%%%%%%%%%%%%%%%%%%%%%%%%%%%%%%%%%%%%%%%%%%%%%%%%
Combining the transformations on the different subspaces we can
write the transformation that diagonalizes $H_{1}^{\text{eff}}$ in
the Fock space as
\begin{equation}\label{S2t}
    R_{2}=R_{2,\bot}+R_{(0,2,-)}+P_{(0,+)}
\end{equation}
At the right hand side of (\ref{apR2}), the three matrices have
entries that commute with each other so we can diagonalize the sum
of them in the atomic Hilbert space (the second step of
$\mathcal{R}_{2,\bot}$ ) as if they had scalar entries. The
eigenvalues of
$R_{2,\bot}^{\dag}H_{1}^{\text{eff}}P_{\bot}R_{2,\bot}$ are thus:

\begin{eqnarray}\label{E2-0}
  E_{1,(n=1,-)}^{\text{eff}}&=&\omega-g,~~~~E_{1,(n=1,+)}^{\text{eff}}=0,~~~~E_{1,(n=2,+)}^{\text{eff}}=0,\nonumber\\
  E_{1,(n\geq3,\pm)}^{\text{eff}} &=& \omega(n-1)+\frac{g}{2}(\sqrt{n-2}-\sqrt{n})
                   \pm\frac{1}{2}\left[\left(-2\omega+g(\sqrt{n-2}+\sqrt{n})\right)^{2}+g^{2}(n-1)\right]^{1/2}.
\end{eqnarray}
\end{widetext}
As it can be seen from (\ref{E2-0}) there are two extra zero
eigenvalues  which have been added  by $R_{2,\bot}$ to the
spectrum of $H_{1}^{\text{eff}}$.

Figs. (\ref{tlsw}-a,b) compare respectively the exact spectrum of
$H$ calculated numerically with the spectrum of
$H_{0}^{\text{eff}}=H_{JC}$ given by (\ref{Ejcm}) and of
$H_{1}^{\text{eff}}$ given by (\ref{E2-0}),(\ref{E1eff}). The
crossings of the exact spectrum are all among the eigenvalues with
different parities. It is found that the spectrum of
$H_{0}^{\text{eff}}$ coincides with the exact one only in the
range of quite small coupling. The spectrum of
$H_{1}^{\text{eff}}$ has been modified with respect to the one of
$H_{0}^{\text{eff}}$  by transforming the encircled crossings
between eigenvalues with the same parity into avoided crossings in
the small $g$ region. This procedure to treat resonances can be
iterated to take into account other resonances appearing at larger
values of $g$.  Figs. (\ref{tlsw}-a,b,c,d,e) show  how the
combination of a one-photon RT and consecutive two-photon RTs lift
the artificial degeneracies (marked by circles) of the effective
Hamiltonians. The successive steps which we have implemented
numerically, transform eigenvalue crossings into avoided
crossings. We observe that these RTs also produce an improvement
of the approximations of the spectrum. Fig. (\ref{tlsw}-f) shows
the effect of a KAM transformation after the fourth two-photon RT
which improves quantitatively the result of Fig. (\ref{tlsw}-e).
The divergence of the KAM transformation close to $g=1$ in Fig.
(\ref{tlsw}-e) is due to the presence of  active resonances at
larger values of $g$.
%%%%%%%%%%%%%%%%%%%%%%%%%%%%%%%%%%%%%%%%%%%%%%%%%%%
\section{\label{strong}effective Hamiltonians in the Strong-Coupling Regime}
In this section we use quantum averaging techniques and RT to
obtain the effective Hamiltonians of (\ref{tls}) in the
strong-coupling regime. We derive a formula that reproduces the
spectrum quite accurately in the whole range of $g$ and for all
energies. We consider the Hamiltonian (\ref{tls}) in the strong
coupling regime $g\gg \omega_{0}>0$ , which suggests to decompose
the Hamiltonian $H$ as
\begin{eqnarray}\label{tls-st}
H&=&H_{0}+V,\nonumber\\
    H_{0}&=&\omega
    (N+1/2)\otimes\openone_{2}+g(a+a^{\dag})\otimes\sigma_{x},\nonumber\\V&=&\frac{\omega_{0}}{2}
    \openone\otimes\sigma_{z}.
\end{eqnarray}
that can be interpreted as the system of a quantized field plus
the coupling term perturbed by the two-level atom. We remark that
in this decomposition, $H_{0}$  contains all the unbounded
operators of the complete model and that the perturbation $V$ is a
bounded operator. In this case
$H_{0}(N,a,a^{\dag},\sigma_{z},\sigma_{x};g)$ is integrable since
we can explicitly transform it into a form involving a regular
function exclusively of the action operators $N,\sigma_{z}$ (given
below in Eq. (\ref{rH1s})). To transform $H_{0}$ to a function of
action operators, first we diagonalize the  term
$g(a+a^{\dag})\otimes\sigma_{x}$ in the atomic Hilbert space by
the transformation (\ref{T}):
\begin{equation}\label{Ts}
T^{\dag}HT=\omega (N+1/2)\otimes \openone_{2}+g(a+a^{\dag})\otimes
\sigma_{z} -\frac{\omega_{0}}{2}\openone\otimes \sigma_{x}.
\end{equation}
 Next we apply a second unitary transformation
\begin{equation}\label{U}
    U= \left(%
\begin{array}{cc}
  e^{-\frac{g}{\omega}(a^{\dag}-a)} & 0 \\
  0 & e^{\frac{g}{\omega}(a^{\dag}-a)} \\
\end{array}%
\right),
\end{equation}
  to transform $\omega (N+1/2)\otimes \openone_{2}+
g(a+a^{\dag})\otimes \sigma_{z} $ into a function of only
$N,\sigma_{z}$ (in this case only of $N$):
\begin{eqnarray}
  H_{1}&:=&U^{\dag}T^{\dag}HTU = \left[\omega
  (N+1/2)- \frac{g^{2}}{\omega}\right]\otimes
  \openone_{2}\nonumber\\
  &-& \frac{\omega_{0}}{2}\left(%
\begin{array}{cc}
  0 &  e^{2\frac{g}{\omega}(a^{\dag}-a)}\\
  e^{-2\frac{g}{\omega}(a^{\dag}-a)} & 0 \\
\end{array}%
\right)
\end{eqnarray}
where use has been made of the commutation relations among  $a$,
$a^{\dag}$, $N$ and the Hausdorff formula:
\begin{equation}\label{BH}
    e^{B}Ce^{-B}=C+[B,C]+\frac{1}{2!}[B,[B,C]]+\cdots.
\end{equation}
We decompose $H_{1}$ as
\begin{eqnarray}\label{rH1s}
H_{1}&=&H_{1}^{(0)}+V_{1}, \nonumber\\
H_{1}^{(0)}&=&U^{\dag}T^{\dag}H_{0}TU=\left[\omega
  (N+1/2)-\frac{g^{2}}{\omega}\right]\otimes
  \openone_{2},\nonumber\\V_{1}&=&-\omega_{0}/2\left(%
\begin{array}{cc}
  0 &  e^{2\frac{g}{\omega}(a^{\dag}-a)}\\
  e^{-2\frac{g}{\omega}(a^{\dag}-a)} & 0 \\
\end{array}%
\right).%
\end{eqnarray}
The effective Hamiltonian of the system for strong-coupling regime
can thus be written as
\begin{equation}\label{strong-eff}
H_{1}^{\text{eff}}=H_{1}^{(0)}+\Pi_{H_{1}^{(0)}}V_{1}
\end{equation}
 The eigenvalues of $H_{1}^{(0)}$ have a two-fold degeneracy for every
value of $n$ as
\begin{equation}\label{E1-0}
E_{1,(n,\pm)}^{(0)}=\omega(n+1/2)-\frac{g^{2}}{\omega}
\end{equation}
The average  of $V_{1}$ relative to $H_{1}^{(0)}$ is thus
\begin{eqnarray}\label{V1bar}
  \Pi_{H_{1}^{(0)}}V_{1} &=& \sum_{n=0}^{\infty}\{|n,+\rangle\langle n,+|V_{1}|n,-\rangle\langle
  n,-|\nonumber\\&+&
  ~|n,-\rangle\langle n,-|V_{1}|n,+\rangle\langle n,+|\}\nonumber \\
   &=& -\frac{\omega_{0}}{2}\sum_{n=0}^{\infty}f_{n}|n\rangle\langle
   n|\otimes\sigma_{x}~,
\end{eqnarray}with
\begin{eqnarray}
  f_{n} &=& \langle n|e^{\frac{-2g}{\omega}(a^{\dag}-a)}|n\rangle=\langle n|e^{\frac{+2g}{\omega}(a^{\dag}-a)}|n\rangle \nonumber\\
   &=& e^{-2g^{2}/\omega^{2}}\langle n|e^{\frac{-2g}{\omega}a^{\dag}}e^{\frac{+2g}{\omega}a}|n\rangle \nonumber\\
   &=&  e^{-2g^{2}/\omega^{2}}\left(\sum_{j=0}^{n}\frac{(-2g/\omega)^{j}}{j!}\sqrt{\frac{n!}{(n-j)!}}\langle n-j|~\right)
   \nonumber\\
   &\times&\left(~\sum_{i=0}^{n}
   \frac{(+2g/\omega)^{i}}{i!}\sqrt{\frac{n!}{(n-i)!}}|n-i\rangle\right)\nonumber\\
   &=& e^{-2g^{2}/\omega^{2}}\sum_{j=0}^{n}\frac{(2g/\omega)^{2j}(-1)^{j}}{(j!)^{2}}\frac{n!}{(n-j)!} \nonumber\\
   &=& e^{-2g^{2}/\omega^{2}}L_{n}\left(\frac{4g^{2}}{\omega^{2}}\right)
\end{eqnarray}
where the $L_{n}$ are the Laguerre polynomials. We remark that in
the limit of large photon number ($n\rightarrow \infty$), $f_{n}$
can be expressed as a zero-order Bessel function
$J_{0}(\frac{4g\sqrt{n}}{\omega})$ \cite{cohen}. $H_{1}$ can  be
reorganized as
\begin{eqnarray}\label{decH1}
H_{1}&=&H_{1}^{\text{eff}}+(V_{1}-\Pi_{H_{1}^{(0)}}V_{1}),\nonumber\\
    H_{1}^{\text{eff}}&=&\left(\omega
  (N+1/2)-\frac{g^{2}}{\omega}\right)\otimes\openone_{2}-\frac{\omega_{0}}{2}F\otimes\sigma_{x},\nonumber\\
(V_{1}-\Pi_{H_{1}^{(0)}}V_{1})&=&-\frac{\omega_{0}}{2}\left(%
\begin{array}{cc}
  0 & G-F\\
  G^{\dag}-F & 0 \\
\end{array}%
\right),
\end{eqnarray}
where
\begin{equation}\label{GF}
    G=e^{\frac{+2g}{\omega}(a^{\dag}-a)},~~~~F=\sum_{n=0}^{\infty}f_{n}|n\rangle\langle n|.
\end{equation}
%%%%%%%%%%%%%%%%%%%%%%%%%%%%%%%%%%%%%%%%%%%%%%%%%%%%%%%%%%%% FIGOR
\begin{figure}
  \includegraphics[width=8.5cm]{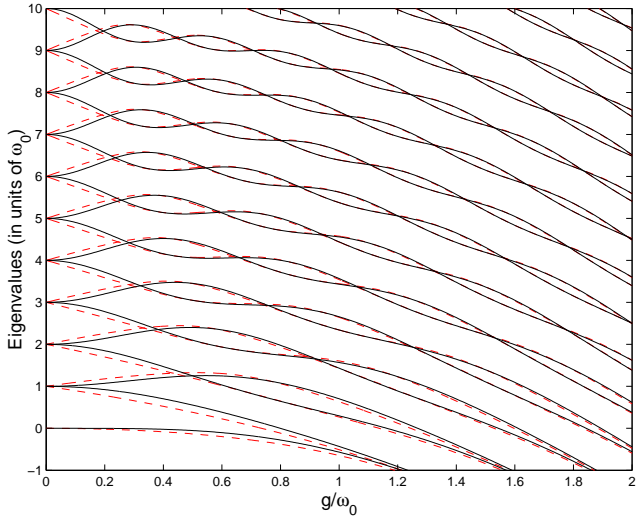}\\
  \caption{Comparison of exact numerical eigenvalues (dashed lines) of (\ref{tls}) as a function of the
  coupling constant in the resonant case ($\omega=\omega_{0}$),
   with the approximate eigenvalues (solid lines) obtained from (\ref{E3-0}).}\label{tlsav}
\end{figure}
%%%%%%%%%%%%%%%%%%%%%%%%%%%%%%%%%%%%%%%%%%%%%%%%%%%%%%%%%%%%%%%%%%%%%
  $H_{1}^{\text{eff}}$ can easily be diagonalized by applying
  the transformation (\ref{T}) that diagonalizes  $\sigma_{x}$:
\begin{equation}
  H_{2}:=T^{\dag}H_{1}T =H_{2}^{(0)}+V_{2},
\end{equation}
with
\begin{eqnarray}
  H_{2}^{(0)} &=& T^{\dag}H_{1}^{\text{eff}}T\nonumber\\
  &=&\left(\omega (N+1/2)-\frac{g^{2}}{\omega}\right)\otimes
  \openone_{2}-\frac{\omega_{0}}{2}F\otimes\sigma_{z},
\end{eqnarray}
and
\begin{eqnarray}\label{V2st}
  V_{2} &=& T^{\dag}(V_{1}-\Pi_{H_{1}^{(0)}}{V}_{1})T\nonumber\\&=&-\omega_{0}/4\left(%
\begin{array}{cc}
  G+G^{\dag}-2F & G-G^{\dag} \\
  -G+G^{\dag} & -G-G^{\dag}+2F \\
\end{array}%
\right).
\end{eqnarray}
 The eigenvalues of $H_{2}^{(0)}$ are therefore
\begin{equation}\label{E3-0}
    E_{2,(n,\pm)}^{(0)}=\omega (n+1/2)-\frac{g^{2}}{\omega}\mp \frac{\omega_{0}}{2}
    e^{-2g^{2}/\omega^{2}}L_{n}(\frac{4g^{2}}{\omega^{2}}).
\end{equation}
which is the same result obtained in \cite{graham1,tur1,franchok}
by other methods. Figure (\ref{tlsav}) compares the exact
numerical spectrum of (\ref{tls}) with the approximation
(\ref{E3-0}) for the resonant case $\omega=\omega_{0}$. One can
see that for large enough $g$, the formula  (\ref{E3-0})
reproduces well the spectrum. It is not very accurate for small
values of $g$ because of the presence of the one-photon zero-field
resonances that we analyze as follows. In the limit $g \rightarrow
0$, we have
\begin{eqnarray}
  H_{2}^{(0),g\rightarrow 0}  &\rightsquigarrow&   \omega (N+1/2)\otimes\openone_{2}-\frac{\omega}{2}\openone
  \otimes\sigma_{z},\nonumber\\V_{2}^{g\rightarrow 0} &\rightsquigarrow& g\left(%
\begin{array}{cc}
  0 & a-a^{\dag} \\
  -(a-a^{\dag}) & 0 \\
\end{array}%
\right).
\end{eqnarray}
Thus degeneracies of $H_{2}^{(0),g\rightarrow 0}$ occur as
\begin{equation}\label{degE3-0}
E_{2,(n,+)}^{(0),g\rightarrow 0}=E_{2,(n-1,-)}^{(0),g\rightarrow
0}.
\end{equation}
They are made active by the resonant terms of $V_{2}^{g\rightarrow
0}$:
\begin{equation}
  V_{2,res}^{g\rightarrow 0}=\Pi_{H_{2}^{(0)}}^{g\rightarrow 0}V_{2}^{g\rightarrow 0} = -g\left(%
\begin{array}{cc}
  0 & a^{\dag} \\
  a & 0 \\
\end{array}%
\right).
\end{equation}
The transformation (the reduction step of the RT) which transforms
this resonant term to a regular function of $N$ is
\begin{equation}\label{zerorwt}
    R_{1}:=\left(%
\begin{array}{cc}
  \openone & 0 \\
  0 & (aa^{\dag})^{-1/2}a \\
\end{array}%
\right)=\left(%
\begin{array}{cc}
  \openone ~~& 0 \\
  0 ~~& \sum_{n=0}^{\infty}|n\rangle\langle n+1| \\
\end{array}%
\right),
\end{equation}
with the properties
\begin{equation}\label{propS1}
R_{1}R^{\dag}_{1}=\openone_{\mathcal{K}},~~~~R^{\dag}_{1}R_{1}=\openone_{\mathcal{K}}-\left(%
\begin{array}{cc}
  0 & 0 \\
  0 & |0\rangle\langle0| \\
\end{array}%
\right),
\end{equation}
% which leads to
%\begin{equation}\label{Sres}
 %   R_{1}^{\dag}V_{2,res}^{g\rightarrow0}R_{1}=-g\sqrt{N}\otimes \sigma_{x}.
%\end{equation}
 We remark that the definition of $R_{1}$ depends on the type of
resonant terms. The reduction step of the RT presented here is
different from (\ref{rwt7}). The Hamiltonian transformed under
this RT has an extra zero eigenvalue corresponding to spurious
 eigenvector $|0,-\rangle$, while for the Hamiltonian transformed
under (\ref{rwt7}), the extra zero eigenvalue corresponds to
$|0,+\rangle$. Applying $R_{1}$ on $H_{2}$ gives
\begin{widetext}
\begin{eqnarray}
  H_{3} &:=& R_{1}^{\dag}H_{2}R_{1}=\left(\omega N-\frac{g^{2}}{\omega}\right)\otimes\openone_{2}+R_{1}^{\dag}V_{2}R_{1}\nonumber\\
   &+&\left(%
\begin{array}{cc}
  \frac{\omega}{2}\left(1-\sum_{n=0}^{\infty}f_{n}|n\rangle\langle
  n|\right) & 0 \\
  0 & -\frac{\omega}{2}\left(1-\sum_{n=1}^{\infty}f_{n-1}|n\rangle\langle
  n|\right)-\left(\frac{\omega}{2}+\frac{g^{2}}{\omega}\right)|0\rangle\langle0| \\
\end{array}%
\right).
\end{eqnarray}
%%%%%%%%%%%%%%%%%%%%%%%%%%%%%%%%%%%%%%%%%%%%%%%%%%%%%%%%%%%%%%%%%%%%%%%%%%%%%%%%%
\begin{figure}
  \includegraphics[width=8.5cm]{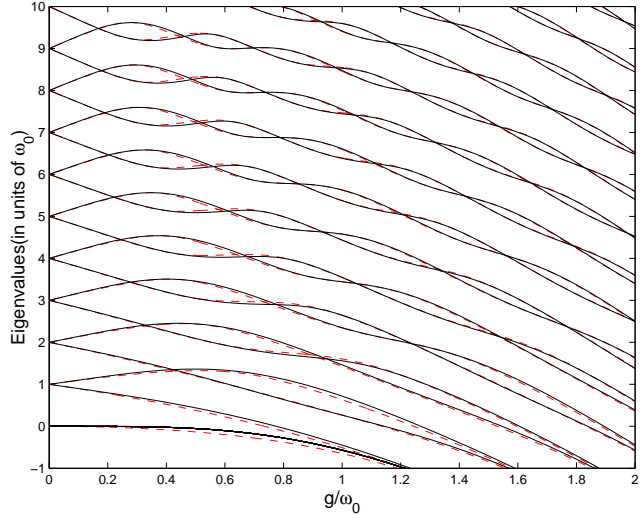}\\
  \caption{Comparison of exact numerical spectrum  of (\ref{tls}) (dashed lines) as a function of the
  coupling constant in the resonant case ($\omega=\omega_{0}$), with  the quite accurate
 result (\ref{E5-0}) which has treated the zero-field resonances by a RT (solid lines).}\label{avRZav}
\end{figure}
%%%%%%%%%%%%%%%%%%%%%%%%%%%%%%%%%%%%%%%%%%%%%%%%%%%%%%%%%%%%%%%%%%%%%%%%%%%%%%%%%%
Next, we take $H_{3}^{(0)}=\omega N\otimes \openone_{2}$ and the
rest of $H_{3}$ as $V_{3}$. Since $H_{3}^{(0)}$ has a two-fold
degeneracy as $E_{3,(n,+)}^{(0)}=E_{3,(n,-)}^{(0)}$, the average
of $V_{3}$ relative to $H_{3}^{(0)}$ is thus
\begin{equation}\label{avV4}
    \Pi_{H_{3}^{(0)}}V_{3}=\left(%
\begin{array}{cc}
  \frac{\omega}{2}-\frac{g^{2}}{\omega}-\frac{\omega}{2}\sum_{n=0}^{\infty}f_{n}|n\rangle\langle
  n|  & \sum_{n=1}^{\infty}-\frac{g}{\sqrt{n}}e^{-2g^{2}/\omega^{2}}L_{n-1}^{(1)}(\frac{4g^{2}}{\omega^{2}})|n\rangle\langle
  n|  \\
  \sum_{n=1}^{\infty}-\frac{g}{\sqrt{n}}e^{-2g^{2}/\omega^{2}}L_{n-1}^{(1)
  }(\frac{4g^{2}}{\omega^{2}})|n\rangle\langle
  n|  & -(\frac{\omega}{2}+\frac{g^{2}}{\omega})(1-|0\rangle\langle0|)+\frac{\omega}{2}\sum_{n=1}^{\infty}f_{n-1}|n\rangle\langle
  n|  \\
\end{array}%
\right),
\end{equation}
where we have used the relation \cite{frasca}
\begin{equation}\label{Lmn}
    \langle
    m|e^{\pm\frac{2g}{\omega}(a^{\dag}-a)}|n\rangle=\sqrt{\frac{n!}{m!}}\left(\frac{\pm2g}{\omega}\right)^
    {m-n}e^{-\frac{2g^{2}}{\omega^{2}}}L_{n}^{(m-n)}\left(\frac{4g^{2}}{\omega^{2}}\right),
\end{equation}
with $L_{n}^{(m-n)}(x)$ the associated Laguerre polynomials. The
new effective Hamiltonian can thus be written as
\begin{equation}\label{H5-0}
H_{3}^{\text{eff}}=\omega N\otimes
\openone_{2}+\Pi_{H_{3}^{(0)}}V_{3}.
\end{equation}
Since all the entries of $H_{3}^{\text{eff}}$ commute with $N$, it
can be diagonalized in the atomic Hilbert space as if its entries
were scalars. The eigenvalues of $H_{3}^{\text{eff}}$ are thus
\begin{eqnarray}\label{E5-0}
  E_{3,(0,-)}^{\text{eff}} &=& 0,~~~~~~~~~~~~E_{3,(0,+)}^{\text{eff}} =
  \frac{\omega}{2}-\frac{g^{2}}{\omega}-\frac{\omega}{2}e^{\frac{-2g^{2}}{\omega^{2}}},
   \nonumber \\
  E_{3,(n\geq 1,\pm)}^{\text{eff}} &=& n\omega-\frac{g^{2}}{\omega}-\frac{\omega}{4}e^{\frac{-2g^{2}}{\omega^{2}}}
  \left(L_{n}(\frac{4g^{2}}{\omega^{2}})-L_{n-1}(\frac{4g^{2}}{\omega^{2}})\right)\nonumber\\
   &\pm& \frac{1}{2}\left[\left(\omega-\frac{\omega}{2}e^{\frac{-2g^{2}}{\omega^{2}}}\left(L_{n}(\frac{4g^{2}}
   {\omega^{2}})+ L_{n-1}(\frac{4g^{2}}{\omega^{2}})\right)\right)^{2}+\frac{4g^{2}}{n}e^{\frac{-4g^{2}}{\omega^{2}}}
   \left(L_{n-1}^{(1)}(\frac{4g^{2}}{\omega^{2}})\right)^{2}\right]^{1/2}.
\end{eqnarray}
\end{widetext}
The zero eigenvalue is the extra spurious one that  has been added
by the RT to the spectrum. Fig. (\ref{avRZav}) compares the exact
numerical spectrum of (\ref{tls}) and the approximation
(\ref{E5-0}) which has treated the zero-field resonances by a
 RT. The figure shows that treating all the active resonances of the
 system allows to obtain all the qualitative features of the spectrum in the whole range
 of the coupling constant and for all energies. At a second stage, since we have treated all the active resonances,
 we can improve further this
 spectrum quantitatively by a KAM-type perturbative iteration.

\section{\label{conc}Conclusions}
We have presented a non-perturbative method based on the quantum
averaging technique to  determine the spectral properties of
systems containing resonances. It consists in the construction of
unitary or isometric transformations that leads to an effective
reduced Hamiltonian. These transformations are composed of two
qualitatively distinct stages. The first one consists of
non-perturbative transformations (RTs) that are adapted to the
structure of the resonances. Their role is to construct a first
effective Hamiltonian that contains the main qualitative features
of the spectrum -- crossings and avoided crossings -- in a given
range of the coupling parameter. The diagonalized form of this
effective Hamiltonian, which depends parametrically  on the
coupling constant, is then taken as a new reference Hamiltonian
around which one can apply perturbative techniques to improve the
quantitative accuracy of the spectrum. We formulate the
perturbative approach in terms of a KAM-type iteration of contact
transformations.  Similar results can be obtained with other
formulations of perturbation theory.

We have illustrated the method with a model of a two level atom
interacting with a single mode of a quantized  field. The method
can be applied to more general systems with several field modes.
It can also be adapted to the treatment of semi-classical models
in which the field is described as a time-dependent function.

We have analyzed the resonances in two regimes of weak and strong
coupling. The results we obtained in the weak-coupling regime can
be expected to be applicable to quite general models. The analysis
of the strong-coupling regime of this model leads to results that
are valid for all values of the coupling and for all energies. The
possibility to obtain such a global result is due to a particular
property of the model, and one cannot expect to obtain it for
general models. The particular property is that the part we
selected as the reference Hamiltonian $H_{0}$ in the
strong-coupling regime contains all the unbounded operators of the
complete model and is explicitly solvable. The term that was left
to be treated by RT and perturbation theory is a bounded operator.

\begin{acknowledgments}
M. A-T wishes to acknowledge the financial support of the French
Society SFERE and the MSRT of Iran. We acknowledge support for
this work from the Conseil R\'{e}gional de Bourgogne.
\end{acknowledgments}
%\section*{Refrences}

\end{document}